\documentclass[preprintnumbers, floatfix, preprintnumbers, letterpaper, superscriptaddress,nofootinbib]{revtex4}
\usepackage{graphicx}
\usepackage{microtype}
\usepackage{amsmath}
\usepackage{amssymb}
\usepackage{subfigure}
\usepackage{hyperref}
\usepackage{url}
\usepackage{xcolor}
\usepackage{color}
\usepackage{mathrsfs}
\usepackage{calrsfs}
\usepackage{amsfonts}
\usepackage{lipsum}
\usepackage{eufrak}
\usepackage{tabularx}
\usepackage{eucal}
\usepackage{latexsym}
\usepackage{ragged2e}
\usepackage{epsfig}
\usepackage{textcomp}

\usepackage{caption}
\DeclareCaptionJustification{justified}{\leftskip=0pt \rightskip=0pt \parfillskip=0pt plus 1fil}
\definecolor{vividviolet}{rgb}{0.62, 0.0, 1.0}
\definecolor{amaranth}{rgb}{0.9, 0.17, 0.31}
\definecolor{palatinateblue}{rgb}{0.15, 0.23, 0.89}
\definecolor{brightpink}{rgb}{1.0, 0.0, 0.5}
\definecolor{cornflowerblue}{rgb}{0.39, 0.58, 0.93}
\definecolor{deepcarminepink}{rgb}{0.94, 0.19, 0.22}
\definecolor{radicalred}{rgb}{1.0, 0.21, 0.37}

\hypersetup{ linktoc=all,
	colorlinks, linkcolor={palatinateblue},
	citecolor={brightpink}, urlcolor={amaranth}
}

\graphicspath{{Images/}}

\renewcommand{\d}[1]{\ensuremath{\operatorname{d}\!{#1}}}

\def\sideremark#1{\ifvmode\leavevmode\fi\vadjust{\vbox to0pt{\vss
			\hbox to 0pt{\hskip\hsize\hskip1em
				\vbox{\hsize1.3cm\tiny\raggedright\pretolerance10000
					\noindent #1\hfill}\hss}\vbox to8pt{\vfil}\vss}}}%
\def\beq{\begin{equation}}
\def\eeq{\end{equation}}

\begin{document}

\title{Understanding Gravitational Entropy of Black Holes: \\ A New Proposal via Curvature Invariants}

\author{Daniele \surname{Gregoris}}
\email{danielegregoris@libero.it}
\affiliation{Department of Physics, School of Science, Jiangsu University of Science and Technology, Zhenjiang 212003, China}

\author{Yen Chin \surname{Ong}}
\email{ycong@yzu.edu.cn}
\affiliation{Center for Gravitation and Cosmology, College of Physical Science and Technology, Yangzhou University, \\180 Siwangting Road, Yangzhou City, Jiangsu Province  225002, China}
\affiliation{Shanghai Frontier Science Center for Gravitational Wave Detection, Shanghai Jiao Tong University, Shanghai 200240, China}

\begin{abstract}

Partly motivated by the arrow of time problem in cosmology and the Weyl curvature hypothesis formulated by Roger Penrose, previous works in the literature have proposed -- among other possibilities -- the square of the Weyl curvature, as being the underlying entropy density function of black hole entropy, but the proposal suffers from a few drawbacks. In this work, we propose a new entropy density function also based solely on the Weyl curvature, but adopting some other combinations of curvature invariants. As an improvement we find that our method works for all static black hole solutions  in four and five dimensional general relativity  regardless of whether they are empty space solutions or not. It should also be possible to generalize our method to higher dimensions. This allows us to discuss the physical interpretation of black hole entropy, which remains somewhat mysterious. Extending to modified theories of gravity, our work also suggests that gravitational entropy in some theories is a manifestation of different physical effects since we need to choose different combinations of  curvature quantities. 
\end{abstract} 

\maketitle

\section{Introduction: The Arrow of Time and Gravitational Entropy}
\label{ssintro}

Time flows from the past to the future, the distinction between the two is that the future has higher entropy. The arrow of time is reflected by the second law of thermodynamics, which is understood in terms of Boltzmann's entropy formula $S=k_B \ln \Omega$, where $k_B$ is the Boltzmann constant and $\Omega$ is the number of available microstates of a given system\footnote{Hereinafter, we will work in the units $c=G=\hbar=k_B=1$.}. Simply put, the statistical mechanics explanation is that there are more ways for the system to be in what we call ``high'' entropy state. For example, gas molecules released in the center of a box tend to fill the box. Physics does not prevent the molecules from coming together by chance and re-gather at the center, it just does not happen in practice because such an event has an extremely low probability. However, probability argument alone has no distinction towards the past or the future, and since physical laws are fundamentally time symmetric, the arrow of time can only be explained by the initial conditions. To quote Feynman \cite{feynman}, ``for some reason, the universe at one time had a very low entropy for its energy content, and since then the entropy has increased...''. Therefore, ultimately this is a problem of cosmology: why was the entropy so low at the Big Bang? (See, however, \cite{1602.05601}.) This problem has received a lot of attention in the literature, see, for example, \cite{Price1, Price2, Price3, vaas, wcc1, penrose, 1108.0417, 0410270, 1305.3836, cyclic, mcinnes1, mcinnes2,1310.5167,2106.15692,2108.07101,2108.10074}.

Therefore, we must first understand what it means for entropy to be low in the very early Universe. The observations of the cosmic microwave background (CMB) indicates that at the time of recombination, matter and radiation (together referred to as the ``matter sector'') has reached thermal equilibrium, as shown by the Planckian black body spectrum. However, the Universe as a whole was \emph{not} in thermal equilibrium, for otherwise it will stay more or less unchanged and no interesting structures can ever emerged, except due to occasional fluctuations \cite{1108.0417}. The low entropy reservoir was in the gravity sector: the gravitational degrees of freedom were not thermalized. To understand this we recall that a uniformly distributed hot gas in a box is in what we ordinarily would refer to as a high entropy state, \emph{in the absence of gravity}. When gravity is present, the natural tendency is for matter to clump due to their mutual attraction. Thus an extremely uniformly distributed hot gas in the very early Universe  (with the density perturbation being mere $\delta \rho/\rho \sim 10^{-5}$)  was of low gravitational entropy -- which has since increased via gravitational collapse. To understand the initial low entropy of the Universe thus amounts to understand why the gravitational entropy was so low at the Big Bang\footnote{One proposal is to weaken gravity in the early Universe by varying the  gravitational constant $G$, so that the uniformly distributed matter did indeed correspond to a ``more typical'' high entropy state \cite{0911.0693}.}. This is a difficult task because the notion of ``gravitational entropy'' -- a term that we have been using with impunity thus far -- is far from being well understood \cite{1012.4476}.

In fact, one should distinguish between entropy caused \emph{by} gravity acting on matter, and entropy \emph{of} gravity -- hereinafter by gravitational entropy we mean the latter. Wallace has argued that gravitational entropy is irrelevant in most contexts except in black hole physics, and that it is enough to consider the dynamics caused by gravitational interactions \cite{0907.0659}. In any case, let us consider gravitational collapse and formation of a black hole thereafter. As we explained above, this process conforms to the second law of thermodynamics. In fact,  black hole formation leads to an enormous increase in entropy. Black holes possess a lot more entropy compared to a typical matter configuration of the same size and energy. Bekenstein-Hawking entropy scales as $S \sim A$, whereas for typical matter configurations, $S \sim A^{3/4}$ \cite{0908.1265v1, hara, 2003.10429}. Consequently, during gravitational collapse of a star, the entropy increases by a factor of $10^{20}(M/M_{\odot})^{1/2}$, where $M_{\odot}$ denotes a solar mass \cite{hara}. Once a black hole is formed, provided there is no electrical charge, then it is now a vacuum solution and thus black hole entropy is purely gravitational. Thus gravitational entropy makes sense at least in the case of black holes. Therefore to properly take into account the evolution of entropy budget in the Universe, we still need to understand entropy \emph{of} gravitational fields. Of course, even if we are not concerned  with the arrow of time or cosmology in general, the problem is itself interesting in black hole physics: what is the physical interpretation of the Bekenstein-Hawking entropy? What does it actually measure\footnote{It is in some sense a measure of our ignorance. As Bekenstein put it \cite{bek}, black hole entropy is ``the measure of the \emph{inaccessibility} [original emphasis] of information (to an exterior observer) as to which particular internal configuration of the black hole is actually realized in a given case.'' See also the recent analysis of the historical development \cite{2102.11209}.  However, see also the objections raised in \cite{1708.05631}.}? 

This question can be asked at two levels: at semi-classical level and quantum gravitational level. An example of the latter is the approach of Strominger and Vafa \cite{9601029}, which demonstrated with a very specific example in string theory that a collection of branes turns into an extremal black hole when a relevant coupling becomes strong, and reproduces the Bekenstein-Hawking entropy. Nevertheless, since the Bekenstein-Hawking entropy is a quantity that is already defined in semi-classical setting, it might be possible -- and even desirable -- to seek a description of the underlying physics at this level by appealing to geometric quantities such as curvatures, regardless of the nature of the more fundamental description. Perhaps by doing so we may glimpse some hints of the microstates.

Penrose introduced the Weyl curvature hypothesis\footnote{Implementing the Weyl curvature hypothesis in a cosmological setting is not a trivial task, and some of its explicit  realizations delivered inconsistent results for physically plausible cosmologies \cite{2004.10222}.} which states that the initial Big Bang singularity should have zero Weyl curvature, whereas singularities in black holes resulted from gravitational collapse and any putative Big Crunch singularity would have large Weyl curvature \cite{wcc1, penrose}. Since the Weyl curvature quantifies tidal deformations, this is just the statement that we expect black hole and Big Crunch singularities to exhibit very messy and chaotic curvature behavior, perhaps like those in the BKL description \cite{BKL,1304.6905}. In contrast, Weyl curvature was zero at the Big Bang, which means that it was much more ``orderly'' in some sense, corresponding to a low entropy state. It turned out that Penrose never meant for the Weyl tensor to be a measure of gravitational entropy in general \cite{9906002}. Nevertheless, motivated by the Weyl Curvature Hypothesis, various proposals for such a measure based on the Weyl curvature have indeed been explored in the literature, see, e.g., \cite{form7,pav}. The simplest of which -- in 5 dimensions -- is to use the squared Weyl tensor as the entropy density function:
\beq
\label{prop}
S_{\rm grav}=\int C_{abcd} C^{abcd} \d V_4\,,
\eeq
where $\d V_4=\sqrt{h}~ \text{d}^4x$ is the hypersurface volume element.
In fact, this procedure reproduces correctly the Bekenstein-Hawking entropy 
\beq
\label{bek}
S_{\rm grav}=\frac{A}{4}=\pi^2 r_H^3
\eeq
for 5-dimensional Schwarzschild black hole and Schwarzschild-anti-de-Sitter black hole up to a constant prefactor \cite{epjcbh1}.
The 5-dimensional restriction is due to the physical dimension of $C_{abcd} C^{abcd}$; in other dimensions in order to reproduce the correct dimension for entropy, one has to take some power $(C_{abcd} C^{abcd})^k$, but $k$ would depend on spacetime dimensions, which seems rather ad hoc and thus not desirable.
In addition, in this proposal it is necessary to remove a region close to the singularity $r=0$,  which otherwise would result in a divergent integral. 

Unfortunately, even if we omit these technical issues, the proposal (\ref{prop}) \emph{cannot} reproduce the gravitational entropy of 5-dimensional Gauss-Bonnet black holes \cite{epjcbh2}, and even that of 5-dimensional Reissner-Nordstr\"om black holes \cite{epjcbh1}. Although it is possible that proposal (\ref{prop}) only works in Einstein's gravity in pure vacuum, it is certainly desirable to have a better proposal that can work in more general settings. 
This is our aim.  In the following we propose to use the Cartan curvature invariants and the Newman-Penrose scalars to construct a gravitational entropy function. The constructive procedure for our result works in the following way. Considering theories governed by the Lagrangian
\beq
{\mathcal L} = \frac{ R}{2}	-\frac{1}{2}g^{ab}\nabla_a \Phi \nabla_b \Phi -V(\Phi)\,,
\eeq
where $\Phi$  is a scalar field and $R$ the Ricci scalar, from \cite[Eq.(96)]{asrev} we can write the entropy as a surface integral over a section of the horizon: 
\beq
\label{walds}
S= \frac{1}{4}\oint \text{d}^2 V \,.
\eeq
This idea was pioneered by Wald  who claimed that black hole entropy is a Noether charge \cite{wald1}, and subsequently re-applied by other authors using the so-called Hamiltonian  method and the first law of thermodynamics (which holds on the horizon) \cite{wald2,corichi,jaco}. We remark that this is a surface integral over the angle variables while the radial coordinate is fixed. This expression is meaningful on an horizon only\footnote{Indeed the location of the horizon depends on the black hole parameters which  govern the thermodynamical evolution of the black hole.}  because if computed on some other arbitrary location $r_a$  it delivers $\pi r_a^2$ which is an irrelevant constant (with respect to the black hole parameters as mass, electric charge, cosmological constant, etc.) playing no role in the thermodynamical description of the black hole  since entropy is a function of state.
Let us now try to apply Gauss theorem to this result. We introduce the inward  and outward  null normals to the horizon $l^\mu$ and $n^\mu$ satisfying the Newman-Penrose normalization $l^\mu n_\mu=-1$ \cite{NPC}. Thus the entropy in (\ref{walds}) is
\beq
S= - \frac{1}{4} \oint \chi_\mu  (l^\mu +n^\mu) \text{d}^2 V\,,
\eeq
where $\chi_\mu$ is an appropriate current to be found. An easy computation reveals that, up to a proportionality constant $\alpha$, the vector we need is $\chi_\mu=l_\mu +n_\mu$. Then, by Gauss theorem we have
\beq
\label{ours}
S= \alpha \int {\rm div} (l +n) \d V\,.
\eeq
In the Newman-Penrose formalism, the divergence of $l^\mu$ is denoted $\rho$ and the divergence of $n^\mu$ is denoted $\mu$ \cite{NPC}, which for static black holes are equal to each other. According to Gauss theorem, only what is inside the boundary is relevant, and therefore the integral in the radial direction should be taken up to the location of the horizon since (\ref{walds}) was computed there (but adding \lq\lq constant" contributions of the type $\pi r_a^2$ (as in \cite{epjcbh1}) would anyway not affect the thermodynamical description).   This is reminiscent of the applicability of the first law of thermodynamics on the horizon, and it is consistent both with the Clifton-Ellis-Tavakol method \cite{form7}, the Bekenstein interpretation that his entropy is a measure of the amount of information hidden by the horizon, and even with the interpretation of  black hole entropy as a quantification of the level of entanglement of the degrees of freedom on the two sides of the horizon since only the boundary between the two subsystem contributes \cite{enta1,enta2}.

 Our paper is intended to  check explicitly proposal (\ref{ours}) in a number of specific configurations, and to point out that thanks to the Bianchi identity it can be used to reconcile the Hawking-Bekenstein entropy formula with the Weyl curvature hypothesis. Actually, due to some algebraic degeneracy of the Weyl tensor in spherically symmetric black hole solutions we will identify various possibilities of choosing the normalization constant. 
 We start with static black holes and show that it works in 4 and 5 dimensions for all static spherically symmetric black hole solutions (not restricted to vacuum). This can be generalized to higher dimensions. Extension to modified theories of gravity, as well as applications to dynamical black holes and cosmology are discussed next. Finally we will conclude with some discussions. In particular, we will also mention the rotating case, with the Kerr solution as a specific example, of which entropy we computed but with some subtle difficulties.

\section{Case I: Static Spherically Symmetric Black Holes}
\label{static}

\subsection{4-dimensional black holes}
\label{static4}

Choosing a system of Schwarzschild-like coordinates, the metric of a static spherically symmetric black hole takes the familiar form
\beq
\label{metric}
\d s^2=   -f(r) \d t^2+\frac{\d r^2}{f(r)}+r^2 \d\theta^2 +r^2 \sin^2 \theta \d\phi^2 \,.
\eeq
In general relativity, explicit solutions admitting these symmetries have been found both in empty spacetime and in matter filled regions, with a wide class of asymptotic behaviors. Being more specific, the Schwarzschild solution represents a case of vacuum and asymptotically flat spacetime, the Reissner-Nordstr\"om black hole is asymptotically flat but supported by an electric field generated by the black hole  charge, the Schwarzschild-(anti) de Sitter black hole arises for a vanishing stress-energy tensor but it is not asymptotically flat due to the effect of a cosmological constant term, and finally the Reissner-Nordstr\"om-(anti) de Sitter spacetime is neither asymptotically flat nor vacuum. In this latter case, the metric function is given by
\beq
\label{frexp}
f(r)=1-\frac{2M}{r}+\frac{Q^2}{r^2}-\frac{\Lambda}{3}r^2\,,
\eeq
where $M$ and $Q$ are respectively the black hole mass and electric charge, and $\Lambda$ the cosmological constant, and the previously mentioned cases can be re-obtained by setting one or two of these parameters to zero. These spacetimes admit one or two horizons located at $r_H$ for which $f(r_H)=0$ depending on the interplay between the values of the various parameters.

For the spacetime (\ref{metric}) we can compute the Newman-Penrose scalar \cite{NPC} $\Psi_2$  and the following component of the first order frame derivative of the Weyl tensor $W=C_{abcd}$ as\footnote{For Petrov type D spacetimes, which include the Kerr-Newman family, $\Psi_2=C_{abcd} n^a m^b \bar{m}^c l^d$ is the only nonzero Weyl scalar \cite{spinorbook}.  Since each frame component of the Weyl tensor and its derivative are Cartan {\it scalar} with physical significance, we prefer to omit the tensorial indices in our notation for simplicity. Furthermore, taking into account the algebraic degeneracy of the Weyl tensor in the spacetimes we are investigating in this paper, which implies that some frame components of the Weyl tensor only differ from each other by a multiplicative constant, we adopt $W$  as a unified notation which can either stand for $\Psi_2$, the frame component $\hat \phi\hat r \hat \theta \hat t$ or $\hat \phi \hat \theta \hat \theta \hat \phi$, etc. (and the same for $DW$).  The reason is that we will derive our results up to a multiplicative constant anyway.   } 
\begin{eqnarray}
\label{weyl1}
\Psi_2=\frac{r^2 f'' -2rf' +2f -2}{12 r^2}\,, \qquad D W=\frac{\sqrt{2 f}(r^2 f'' -2rf' +2f -2)}{8 r^3}\,,
\end{eqnarray}
where a prime denotes a derivative with respect to the radial coordinate $r$, and $D \equiv n^a \nabla_a$ is the namesake Nemwan-Penrose directional derivative. 
These results have been found adopting the null coframe
\beq
\label{coframe}
l_a=\frac{1}{\sqrt{2}}\left(\sqrt{f(r)}\d t -\frac{\d r}{\sqrt{f(r)}} \right)\,, \qquad n_a=\frac{1}{\sqrt{2}}\left(\sqrt{f(r)}\d t +\frac{\d r}{\sqrt{f(r)}} \right)\,, \qquad m_a=\frac{r \d\theta +i \,r \sin \theta \d\phi}{\sqrt{2}}   \,,
\eeq
where $i^2=-1$, which allows us to cast the spacetime metric (\ref{metric}) as
\beq
\d s^2=-2 l_{(a}n_{b)}+2 m_{(a}\bar m_{b)}\,,
\eeq
with an overbar denoting a complex conjugation, and round parentheses standing for symmetrization. The null coframe is characterized by the relations $l_a l^a=n_a n^a=m_a m^a =\bar m_a \bar m^a=0$ and $ -l_a n^a=1=m_a \bar m^a$. This is the ``canonical frame'' for the metric (\ref{metric}) \cite{advgr} making the quantity $DW$  a Cartan curvature invariant with the property of being foliation-independent \cite{cartanref}.

Unlike the aforementioned proposal (\ref{prop}) which requires taming the divergence at the pole $r=0$ by an arbitrary cutoff,  by considering (\ref{weyl1}) we can do away with the need of regularization and simply define gravitational entropy by
\beq
\label{entropyformula}
S_{\rm grav}:=\frac{1}{3 \sqrt{2}}\int_0^{r_H} \int_\Omega \Big| \frac{D W}{\Psi_2}\Big| \frac{r^2 \sin \theta }{\sqrt{f(r)}} \d r \d\theta \d\phi =\pi r_H^2=\frac{A}{4}\,,
\eeq
in agreement with the entropy-area law, or in other words with the holographic principle \cite{hoft,holo2}. The proportionality constant $1/(3\sqrt{2})$ will be discussed in subsec. (\ref{5d}).
The integration over $r$ is somewhat mysterious as $r$ is a temporal coordinate inside the black hole. For the Schwarzschild and Schwarzschid-(A)dS cases, one might be tempted to explain this by arguing that entropy has to do with time evolution, and so it is natural to integrate over a timelike coordinate. In fact, Edery and Constantineau have proposed to identify gravitational entropy with the nonstationary nature of black hole interior \cite{1010.5844} for essentially the same reason. However, if this proposal is correct, then extremal black holes would have zero entropy, which goes against the current consensus (especially in the string theory and holography community) that extremal black holes should also have entropy $A/4$ (this is not a settled debate, however, see \cite{0901.0931,1502.02737}). Furthermore, our approach of integrating over $r \in (0,r_H)$ correctly reproduces the Bekenstein-Hawking entropy of a Reissner-Nordstr\"om black hole regardless of whether $\partial/\partial r$ is timelike or spacelike in the interval. Thus it would seem that gravitational entropy defined in this way is not directly related to stationarity of the interior region.  The following point should be noted: although the integration up to any arbitrary $r$ of our entropy density delivers a geometrical result which is an area, this quantity bears the physical meaning of an entropy only if computed at the horizon. We can elucidate this fact by considering the simplest case of a Schwarzschild black hole for which $r_H=2M$: a variation of the black hole entropy $\delta S_{\rm grav}$ would require a variation of the black hole size  $\delta r_H$ which ultimately would require an evolution $\delta M$ of the black hole  inside the parameters phase space. On the other hand a variation of a generic $r$ does not imply a change in the physical properties of the black hole; thus our integral, if computed in some exterior regions,  does not deliver an entropy, consistently with Hawking-Bekenstein interpretation.      

Taking into account that according to the geodesic deviation equation the strength of tidal forces is related to the derivative of the Weyl tensor (see, e.g., \cite{tidal1,tidal2} for some recent investigations along this line in the Reissner-Nordstr\"om and Kiselev spacetimes), we can see that the density of gravitational entropy is directly connected to the strength of tidal forces which are indeed necessary for the existence of a compact object to which an entropy is assigned like the black hole. It is worth noting that tidal forces have been claimed to be a form of gravitational waves \cite{ellis}, although the former depend only on the electric part of the Weyl curvature while the latter require also some gravitomagnetic effects. 

Furthermore, adopting the same language of  Clifton-Ellis-Tavakol \cite{form7} by writing
\beq
S_{\rm grav}= \int_V \frac{\rho_{\rm grav}}{T_{\rm grav}}\d V \,,
\eeq
we can interpret the Cartan invariants $DW$ and $W=|\Psi_2|$  to be the \lq\lq energy density and temperature of the gravitational field", respectively. It should be appreciated that our procedure is not sensitive to the matter content of the spacetime, its asymptotic flatness properties, or to the fact that $f(r)$ should be found by integrating the Einstein equations for having a physically relevant spacetime -- all we require is the existence of a horizon. We also note that the vanishing of the Cartan invariant $DW$ on the horizon, which made it an appropriate quantity for taming the teleological nature of black hole spacetimes \cite{cartan},  is cured by the same property of the function $1/\sqrt{f(r)}$ entering the hyperspace volume element $\d V_3=({r^2 \sin \theta}/{\sqrt{f(r)}}) \d r \d\theta \d\phi$.

In the cosmological contexts investigated in \cite{nicos}, in order to deal with the the so-called ``isotropic singularity'', a rescaling of the Weyl curvature in the gravitational entropy proposal was introduced:
\beq
C_{abcd}C^{abcd} \qquad \longmapsto \qquad \frac{C_{abcd} C^{abcd}}{R_{ab} R^{ab}}.
\eeq
Remarkably, our recipe essentially involves only the Weyl curvature, making it more preferable as a measure of the gravitational entropy density because the matter content of the spacetime does not directly affect it. 

Our result can be extended beyond the metric assumption of (\ref{metric}). In fact, let us now consider the metric of a static spherically symmetric but \emph{deformed} black hole \cite{yunes,yunes2}
\beq
\label{dist}
\d s^2=-f(r)[1+h(r)] \d t^2 +\frac{[1+h(r)] \d r^2}{f(r)} + r^2 \d\Omega^2\,,
\eeq
where in the most general case the ``deformation function'' is written as
\beq
h(r)=\sum_{k=0}^{\infty} \epsilon_k \left( \frac{M}{r}\right)^k\,,
\eeq
with $\epsilon_k$ being the deformation parameters. The Cauchy, Killing and apparent horizon(s) \cite{asrev}  coincide and can still be found by solving $f(r_H)=0$. This framework allows one to study post-Newtonian effects in astrophysical phenomena \cite{ppn1,ppn2,ppn3,ppn4}. Replacing $l_a$ and  $n_a$ in the  null coframe (\ref{coframe}) with
\beq
l_a=\frac{1}{\sqrt{2}}\left(\sqrt{f(r) [1+h(r)]}\d t -\sqrt{\frac{1+h(r)}{f(r)}}\d r \right)\,, \qquad n_a=\frac{1}{\sqrt{2}}\left(\sqrt{f(r) [1+h(r)]}\d t + \sqrt{\frac{1+h(r)}{f(r)}}\d r \right)\,,
\eeq
without modifying $m^a$ and $\bar m^a$ (since the angular part of the metric remains unchanged), we get for the Weyl curvature and its frame derivative:
\begin{eqnarray}
\Psi_2 &=&  \frac{r^2 (1+h)^2 f''+r^2 f(1+h) h''+r(1+h)(r h'-2h-2)f'-f (h')^2 r^2+2(1+h)^2 (f-1-h)}{12 (1+h)^3 r^2} \,, \\
D W &=&    \frac{[r^2 (1+h)^2 f''+r^2 f (1+h) h''+r (1+h) (r h'-2 h-2)f'-(h')^2 f r^2+2 (1+h)^2 (f-1-h)]\sqrt{2f}}{8 (1+h)^{7/2} r^3}   \,,
\end{eqnarray}
which allow us to reproduce the result in (\ref{entropyformula}) simply by noticing that now we should modify the volume element as $\d V_3=\sqrt {\frac{1+h(r)}{f(r)} }r^2 \sin\theta \d r \d\theta \d\phi$ for taking into account the black hole deformations.

 We would like to mention that if we consider a metric tensor obeying the symmetry \cite{0707.3222}
\beq
\label{stellar}
\d s^2 = -A(r) \d t^2 + \frac{\d r^2}{B(r)}+r^2 (\d \theta^2 +\sin^2 \d \phi^2)	\,, \qquad A(r) \neq B(r)\,,
\eeq
and work in the coframe
\beq
l_a=\frac{1}{\sqrt{2}}\left(\sqrt{A(r)}\d t -\frac{1}{\sqrt{B(r)}}\d r \right)\,, \qquad n_a=\frac{1}{\sqrt{2}}\left(\sqrt{A(r)}\d t +\frac{1}{\sqrt{B(r)}}\d r \right)\,, 
\eeq
we can compute 
\beq
\label{psistar}
\Psi_2= \frac{ (2 A A'' B   + A B' A' - (A')^2 B) r^2 -2 A (A B'  +  A' B) r + 4 A^2 (B-1)}{24 (r A)^2}.
\eeq
Therefore a postulated entropy density of the form $\frac{D \Psi_2}{\Psi_2} \equiv \frac{\sqrt{B(r)} \partial_r \Psi_2}{\sqrt{2}\Psi_2}$  is not consistent with the area law. Indeed several \emph{non-black hole} solutions of the Einstein equations which are interpreted as fluid or gaseous sphere in equilibrium (the interior Schwarzschild solution being the prototype), for which the Hawking-Bekenstein law does not hold, come in the form of (\ref{stellar})  \cite[Sect. 16.1]{stephani}. This ``failure'' is consistent with the known results in the literature allowing us to argue that our proposal of entropy density is general enough to cover the relevant cases but not {\it too} general. In other words this example shows that our proposal neither constitutes a trivial numerical coincidence nor a general property of the Weyl tensor; instead it suggests a connection between the physical interpretation of the quantity $DW/\Psi_2$ as an entropy density. 

As an explicit example, the Tolman metric IV comes with\footnote{Should we consider the interior Schwarzschild solution for which $A(r)= \left(X-Y\sqrt{1-\frac{r^2}{R^2}} \right)^2$ and $B(r)= 1-\frac{r^2}{R^2}   $ we would get the trivial result $\Psi_2=0$.}
\beq
A(r) = Y^2 \left( 1+ \frac{r^2}{X^2} \right)\,, \qquad B(r)= \frac{\left( 1-\frac{r^2}{R^2}  \right) \left( 1+\frac{r^2}{X^2}\right)}{1+\frac{2 r^2}{X^2}}\,,
\eeq
with $X$, $Y$ and $R$ being some constants. It describes a sphere of compressible fluid in hydrostatic equilibrium supported by a pressure vanishing at the boundary of the configuration \cite{tov1}.
Specifying (\ref{psistar}) we obtain
\beq
\Psi_2= -\frac{(X^2 +2R^2)r^2}{3R^2(X^2 +2r^2)^2}
\eeq
and
\beq
\int_0^R \Big |  \frac{2(X^2 -2r^2)}{r(X^2+2r^2)}\Big| r^2 \d r  =    \Big[ r^2 -X^2 \ln(X^2 +2r^2)   \Big]\Big|^R_0 =     R^2 -X^2 \ln \left( 1+2\left( \frac{R}{X}\right)^2 \right)<R^2\,.
\eeq
Thus, our method shows that the entropy within this stellar configuration is smaller than what it would be for a black hole. This is consistent with the increase of entropy during black hole formation phase \cite{hara}. 

Now let us remark on the stationary but not static case, with the Kerr solution as an example.
The Kerr black hole is axially symmetric but not spherically symmetric. We can apply the Newman-Janis trick \cite{janis1,janis2} by introducing the complex coordinate $r':= r - ia \cos\theta$ and its complex conjugate $\bar r'$, where $a$ is the rotation parameter. Then we can deal with the resulting line element
\beq
\d s^2 = - \left[ 1-M \left(\frac{1}{r'} +\frac{1}{\bar r'}\right)  \right] \d t^2 + \frac{r' \bar r' \d r'^2}{r' \bar r' -M (r' +\bar r')} + r' \bar r'(\d \theta^2 + \sin^2 \theta \d \phi^2)\,.
\eeq
Hence, {\it mutatis mutandis} we have $\Psi_2 = - \frac{M}{r'^3}$ and $ \frac{D \Psi_2}{\Psi_2} $ is the appropriate entropy density for obtaining $ S_{\rm grav} =\pi r_H' \bar r_H'=\pi (r_H^2 +a^2 \cos^2 \theta_a)$. Unfortunately, this result depends on the angle $\theta$, and only recovers the correct result for the entropy $S=\pi (r_H^2 +a^2)$ if $\theta_a=0$ or $\theta_a=\pi$, i.e., along the axis of rotation.  The problem may be due to mathematical subtleties in the procedures that require further investigations. For example, if tackled by adopting Boyer-Lindquist coordinates, the integrals over $r$ and $\theta$ do not factor and they are obstructed by the black hole singularity which is related by a condition which involves both of these coordinates (e.g.  when we integrate over $r$ how do we take care of the singularity  at $\theta=\pi/2$, if we need to integrate over the angle later on?). We will leave this issue for future investigations.

\subsection{5-dimensional black holes}\label{5d}

Now we will consider the case of the five-dimensional counterpart of (\ref{metric}) by writing
\beq
\label{5dmetric}
\d s^2=   -f(r) \d t^2+\frac{\d r^2}{f(r)}+r^2( \d\theta^2 + \sin^2 \theta \d\phi^2 +  \sin^2\theta \sin^2 \phi \d\omega^2 ) \,,
\eeq
where we can consider\footnote{For convenience, here $M$ and $Q$ are the normalized mass and charge, which are proportional to the ADM one, so we can avoid writing some factors of $\pi$ and so on.} 
\beq
f(r)=1-\frac{2M}{r^2} +\frac{Q^2}{r^4}-\frac{\Lambda r^2}{6}
\eeq
for an explicit solution which -- for the $\Lambda<0$ case -- is of interest in string theory and for applications of the AdS/CFT correspondence \cite{5dref}. By following \cite{cartan} [Eqs.(4.1.2)-(4.1.3)] we introduce the following canonical frame
\beq
l_a=\frac{1}{\sqrt{2}}\left( \frac{\d r}{\sqrt{f(r)}} -\sqrt{f(r)} \d t  \right), \quad n_a=\frac{1}{\sqrt{2}}\left( \frac{\d r}{\sqrt{f(r)}} + \sqrt{f(r)} \d t  \right), \quad m_{a1}=r \d \theta , \quad  m_{a2}=r \sin \theta \d \phi , \quad m_{a3}=r \sin\theta \sin\phi \d \omega , 
\eeq
in terms of which the metric (\ref{5dmetric}) reads $\d s^2= -2l_{(a} n_{b)} + \Sigma_{i=1}^3   m_{i(a} m_{ b)i}$, and we get
\beq
\label{w5d}
W=\alpha \frac{r^2 f'' -2rf' +2f -2}{ r^2}\,, \qquad DW=\beta \frac{\sqrt{2 f}(r^2 f'' -2rf' +2f -2)}{ r^3}\,.
\eeq
Note that there is a subtle difference here compared to the previous cases: in 5-dimensions we cannot construct a \emph{null} coframe, and so we do not have the Newman-Penrose $\Psi_2$. However we can still explicitly fix a frame and compute the Weyl curvature $W$ and its frame derivative $D W$. 

We can choose either $\alpha={1}/{4}, \, {1}/{12}$ or $\beta={1}/{6}, \, {1}/{12}$. This is because some of the frame components of the Weyl tensor and of its first derivative are algebraically dependent among each other. By the theorems due to Cartan these are scalars (because they have been computed in the frame constructed in \cite{cartan}) and thus they are well-defined integrands. This means that we have some freedom in tuning the constant appearing in front of the integral (if it has some relevant physical meaning) depending on which choices we make. Thus, by recalling that in five dimensions the volume element is $\d V_4= r^3 \sin^2 \theta \sin \phi \d r \d\theta \d\phi \d\omega / \sqrt{f(r)}$, we can obtain the higher-dimensional entropy-area law again as
\beq
\label{entropy5d}
S_{\rm grav} =\gamma  \int_0^{r_H} \int_\Omega \Big | \frac{DW}{W} \Big| \d V_4 = \frac{\pi^2}{2} r^3_H=\frac{A}{4}\,,
\eeq
where we report in Table \ref{tablegamma} the values that $\gamma$ should take according to the choices we make for the values of $\alpha$ and $\beta$. We would like to mention that also in  4-dimensions we have a second possibility if we choose to work with $W=2 \Psi_2$ rather than with $W=\Psi_2$ which will affect the proportionality constant in (\ref{entropyformula}) by a factor 2. 

Similar to the proposal in \cite{epjcbh1}, the proportionality constant $\gamma$ can only be fixed \emph{a posteriori} if we want to get the $1/4$ factor in the area law. In other words, our proposal can only obtain the Bekenstein-Hawking entropy up to a constant prefactor.
However, it should be emphasized that unlike in the analysis in \cite{epjcbh1} we are not restricted to the zero-charge configuration. In addition our proposal should work in higher dimensions, \emph{mutatis mutandis}.

\begin{table}[ht]
	\begin{center}
		\scalebox{1.7}{%
		\begin{tabular}{|c|c|c|}
			\hline
			$\alpha $ &  $\beta$ & $\gamma$   \\
			\hline
			$ \frac{1}{4} $    & $\frac{1}{6} $ &  $\frac{9 \sqrt{2} }{16}$  \\
			\hline
		   $ \frac{1}{4} $    & $\frac{1}{12} $ &  $\frac{9 \sqrt{2} }{8} $  \\
		   \hline
		   $ \frac{1}{12} $    & $\frac{1}{6} $ &  $\frac{3 \sqrt{2} }{16}$  \\
		   \hline
		   $ \frac{1}{12} $    & $\frac{1}{12} $ &  $\frac{3 \sqrt{2} }{8}  $  \\
			\hline
		\end{tabular}}
		\caption{The value that the proportionality constant $\gamma=\frac{3 \sqrt{2} \alpha}{8 \beta}$ should take in the entropy formula (\ref{entropy5d}) depending on the values of $\alpha$ and $\beta$ we consider in (\ref{w5d}).}
		\label{tablegamma}
	\end{center}
\end{table}

\subsection{Extensions beyond general relativity}

Our method of constructing an appropriate entropy density for the Bekenstein area law for static black holes in four and five spacetime dimensions does not need to assume any specific choice for the function $f(r)$ entering the metric tensors (\ref{metric}) and (\ref{5dmetric}): the solution does not need to be vacuum\footnote{For example, it can also be applied to regular black holes, such as the generalized Bardeen solution in the presence of nonlinear electrodynamics \cite{bardeen1,bardeen2,bardeen3}, whose metric function in 4 dimensions is given by
\beq
f(r)=1-\frac{2M r^2}{(r^2 +Q^2)^{3/2}}+\frac{Q^2 r^2}{(r^2 +Q^2)^2}.
\eeq
}, and furthermore it does not need to be a solution of the Einstein's field equations.  Therefore, an area law is reproduced also for black holes arising in modified gravity theories because our method is purely geometrical. However, this may not correspond to an entropy formula because modifications of the underlying gravitational theory affects the Bekenstein law in nontrivial ways (so that in general it is no longer $1/4$ times the horizon area).
To put it differently, black hole entropy is a dynamical quantity (which depends on the field equations), unlike its temperature which is kinematical \cite{9712016,0106111,1011.5593}.
Two examples of modified gravity in which black hole entropy takes a different form ($S \neq A/4$) are $f(R)$ gravity \cite{entropyf1,entropyf2,entropyf3,entropyf4,entropyf5} and the Einstein-Gauss-Bonnet gravity \cite{gbentropy1,gbentropy2,gbentropy3}.

For example, in the Einstein-Gauss-Bonnet case the 5-dimensional black hole has metric function
\beq
f(r)= \sqrt{k +\frac{r^2}{4 \delta} \left( 1-\sqrt{1+\frac{128 \pi\delta M}{3 \Sigma_k r^4} +\frac{4 \delta \Lambda}{3}} \right)},
\eeq
where $\Sigma_k$ is the unit volume of the  manifold with constant sectional curvature $k \in \left\{-1,\,0,\,1\right\}$ (with the former two choices being admissible only if $\Lambda<0$), and $\delta$ denotes the Gauss-Bonnet coupling parameter. The black hole entropy is given by
\beq
S_{\rm grav}= \frac{\pi^2 r_H^3}{2} \left(1+\frac{12 \delta k}{r^2_H}\right).
\eeq

Therefore, our method can be applied directly in the particular case $k=0$, but otherwise some curvature corrections would be required to obtain the second term. Thus far we have not found a good way to achieve this without resorting to some very nontrivial manipulations that do not seem to be as natural. Regardless of whether a better prescription can be found, it is clear that whenever the area law is modified, our method no longer works straightforwardly. This is exactly because our prescription is purely geometrical and does not rely on the field equations. This might hint at a deeper nature of gravitational entropy: since we need to pick different combinations of the curvature objects to account for the correction term to the Bekenstein area law, it is possible that gravitational entropy in different theories of gravity is actually a manifestation of different physical effects (e.g. the Weyl curvature is related to tidal forces). See Sect.\ref{discuss}  for more discussions.

On the other hand, by studying the first law of thermodynamics and using the fact that temperature is the conjugate variable of the entropy, it was argued that the Bekenstein area law still holds in Weyl gravity \cite{bambi} and massive gravity \cite{cai}. Thus, our result for the entropy density would hold for the class of black holes that accounts for Modified Newtonian Dynamics effects via a linear term in the redshift function
\beq
f(r)=1-\frac{2 M}{r} +\varepsilon r- \frac{\Lambda r^2}{3}\,,
\eeq
which arises in general relativity (e.g., the Kiselev solution), conformal gravity and massive gravity \cite{deg1,deg2}.

\section{Dynamical black holes and cosmological applications}

\subsection{General case}

Next, we shall explore dynamical spacetimes. First, let us consider
 a spacetime which can be adopted as a mathematical model for describing  either the collapse of cosmic material leading to the formation of a black hole or an inhomogeneous expanding universe.  The line element of a dynamical spherically symmetric spacetime is given by \cite[p.251]{stephani}
\beq
\label{generaldyn}
\d s^2 = -e^{2 \nu(t,r)} \d t^2 + e^{2\lambda(t,r)}\d r^2 +X^2(t,r) (\d \theta^2 +\sin^2\theta \d \phi^2)\,,
\eeq
where the functions $\nu(t,r)$, $\lambda(t,r)$ and $X(t,r)$ should be found as solutions of the field equations once the matter content of the spacetime is given. By noticing that for static spherically symmetric spacetimes, as the ones investigated in Sect.\ref{static4}, we have $\mu=\rho \propto {D\Psi_2}/{ \Psi_2}$ for the Newman-Penrose spin coefficients, we can construct an appropriate entropy density formula for the spacetime (\ref{generaldyn}) as follows. We introduce the coframe 
\beq
l_a=\frac{e^{ \nu(t,r)}\d t - e^{\lambda(t,r)} \d r }{\sqrt{2}}\,, \qquad n_a=\frac{e^{ \nu(t,r)}\d t + e^{\lambda(t,r)} \d r }{\sqrt{2}}\,, \qquad m_a=\frac{X(t,r) \d\theta +i \,X(t,r) \sin \theta \d\phi}{\sqrt{2}}   \,,
\eeq
and compute 
\beq
\rho = \frac{X'(t,r) e^{-\lambda(t,r)} +\dot X(t,r) e^{-\nu(t,r)}}{\sqrt{2} X(t,r)}\,, \qquad \mu = \frac{X'(t,r) e^{-\lambda(t,r)} - \dot X(t,r) e^{-\nu(t,r)}}{\sqrt{2} X(t,r)}\,,
\eeq
where an overdot stands  for a time derivative. Then, considering the spatial volume element $\d V_3=e^{\lambda(t,r)} X^2(t,r) \sin\theta \d r \d \theta \d \phi$,  the area law in terms of the areal radius $\tilde r :=X(t,r)$ is found as
\beq
S_{\rm grav}= \gamma \int_0^{r_H} \int_\Omega \Big| \frac{\mu +\rho}{2}   \Big| \, \d V_3= \sqrt{2} \pi \gamma \int_0^{r_H} X'(t,r) X(t,r) dr = \frac{\pi \gamma}{\sqrt{2}} X^2(t,r_H) = \pi \tilde r_H^2\,,
\eeq
where in the last step we have chosen $\gamma=\sqrt{2}$.

\subsection{Specializing to the Lema\^itre-Tolman-Bondi spacetime}

A specific subcase of (\ref{generaldyn}) for dust-filled spacetimes is the Lema\^itre-Tolman-Bondi metric \cite{snyder}
\beq
\d s^2=-\d t^2 + \frac{[X'(t,r)]^2}{ 1 +2 E(r)}\d r^2 +X^2(t,r) \d\Omega^2\,.
\eeq
Depending on the sign of the expansion rate, this metric can also be adopted for describing an expanding inhomogeneous universe by  interpreting the function $E(r)$ as the generalization of the constant spatial curvature parameter $k$ of the Friedman universe \cite{enqv}. 
Working now in the coframe
\beq
\label{coframe}
l_a=\frac{1}{\sqrt{2}}\left(\d t -\frac{ X'(t,r) \d r}{\sqrt{1+2E(r)}} \right)\,, \qquad n_a=\frac{1}{\sqrt{2}}\left(\d t +\frac{X'(t,r) \d r}{\sqrt{1+2E(r)}} \right)\,, \qquad m_a=\frac{X(t,r) \d\theta +i \, X(t,r) \sin \theta \d\phi}{\sqrt{2}}  \,,
\eeq 
we obtain
\begin{eqnarray}
\Psi_2 &=& \frac{-X^2 \ddot X'+X \dot X \dot X'+X \ddot X X'+(2E-\dot X^2) X'-X E'}{6 X^2 X'} \,, \\
DW &=& \frac{\sqrt{2} [-X^2 \ddot X'+X \dot X \dot X'+X \ddot X X'+(2E-\dot X^2) X'-X E'](\sqrt{1+2E} +\dot X)}{4 X^3 X'}    \,,
\end{eqnarray}
for the Weyl scalar and the  frame derivative of the Weyl tensor. Thus:
\beq
 \frac{D W}{\Psi_2}=\frac{3 ( \sqrt{1+2E} +\dot X )}{ \sqrt{2}X}\,.
\eeq 
If we further assume the model to describe a quickly expanding universe, or that the black hole formation process is occurring slowing as compared to the curvature effects, so that $\sqrt{1+2E} \gg - \dot X$, we can again reproduce the Bekenstein law in terms of the areal radius because
\beq
S_{\rm grav} =\frac{1}{3 \sqrt{2}}\int_0^{r_H} \int_\Omega \Big| \frac{D W}{\Psi_2}\Big| \frac{X' X^2 \sin \theta }{\sqrt{1+2E}} \d r \d\theta \d\phi \approx  2 \pi \int_0^{r_H} X(t,r) X'(t,r)dr= \pi X^2(t,r_H)=  \pi \tilde r_H^2\,.
\eeq
This route corresponds to consider the entropy density as given by the Newman-Penrose spin coefficient $\mu$. On the other hand, we get $\rho=\frac{\sqrt{1+2E} -\dot X}{\sqrt{2} X(t,r)}$, which can be used as an entropy density function in the regime in which $\sqrt{1+2E} \gg\dot X$. In the intermediate cases, the appropriate entropy density function to consider is $\propto |\mu +\rho|$, as already discussed in the general framework. As a mathematical remark, we need to mention that the spin coefficients are suitable integrands being scalars in the spacetime considered in this section \cite{coley}.

\subsection{Friedman-Lema\^{i}tre-Robertson-Walker and (Anti-)de Sitter limit}

We have noted that in the case of static and spherically symmetric black holes discussed in Sect. \ref{static4}, the entropy density can be equivalently written in terms of the Newman-Penrose spin coefficients because
\beq
\mu=\rho \propto \frac{D W}{W}.
\eeq
This observation has helped us when finding the entropy density function in dynamical spherically symmetric spacetimes and it also constitutes a hint for extending our formalism to the conformally flat case of the Friedman-Lema\^{i}tre-Robertson-Walker universe. In this case, the metric tensor is
\beq
\d s^2=-\d t^2 +\frac{a^2(t)}{1-kr^2}\d r^2 + a^2(t) r^2 (\d\theta^2 + \sin^2 \theta \d\phi^2)\,,
\eeq
where $k \in \left\{1,\,0,\,-1\right\}$ for the cases of closed, flat, and hyperbolic universe, respectively. 
In the null coframe
\beq
l_a=\frac{1}{\sqrt{2}}\left(\d t -\frac{a(t) \d r}{\sqrt{1-k r^2}} \right)\,, \qquad n_a=\frac{1}{\sqrt{2}}\left(\d t +\frac{a(t) \d r}{\sqrt{1-k r^2}} \right)\,, \qquad m_a=a(t)\frac{r \d\theta +i \,r \sin \theta \d\phi}{\sqrt{2}}, 
\eeq
we get the following result for the Newman-Penrose spin coefficients of interest:
\beq
\rho =  \frac {\sqrt{1-kr^2} -\dot a r}  {\sqrt{2} a(t) r}\,,  \qquad
\mu =  \frac {\sqrt{1-kr^2} +\dot a r}   {\sqrt{2} a(t) r}     \,.
\eeq
Considering the volume element $\d V_3 = \frac{a^3(t) r^2 \sin\theta \d r \d\theta \d\phi}{\sqrt{1-k r^2}}$, and taking into account that for a Friedman-Lema\^{i}tre-Robertson-Walker spacetime the spin coefficients $\rho$ and $\mu$ can be written in terms of scalar quantities \cite{coley} making them suitable as integrand functions,  we can compute\footnote{ Here we have in mind the dynamical apparent horizon defined by the condition $|| \nabla_a \tilde r||^2_H=0$, where $\tilde r := a(t) r$ is the areal radius  \cite{asrev}.}
\beq
S_{\rm grav} = \frac{1}{\sqrt{2}}\int_0^{r_H} \int_\Omega \Big | \frac{\rho +\mu}{2} \Big| \d V_3 = 2 \pi a^2(t) \int_0^{r_H} r \d r = \pi \tilde r_H^2\,,
\eeq
where once again we have introduced the areal radius $\tilde r := a(t) r$.  It should be noted once again that our procedure provides an appropriate entropy density for the Hawking-Bekenstein law only if the upper endpoint of the integral is given by the horizon location. For example for the de Sitter universe we have $\tilde r_H= \sqrt{3/ \Lambda}$, and therefore a variation $\delta S_{\rm grav}$ would require a variation $\delta \tilde r_H$ and then a variation of $\delta \Lambda$, e.g. of the physical properties of the universe. On the other hand a variation $\delta r$ does not require any variation of the physical state of the universe, nor it implies a change in its thermodynamical properties, as indeed the Hawking-Bekenstein entropy is not associated to arbitrary spatial domains but only to those bound by a horizon (in the cosmological case it is better known as the ``Gibbons-Hawking'' entropy \cite{Gibbons}).

\section{Discussion and Open Questions}\label{discuss}

In this work we have proposed a new gravitational entropy density function based \emph{only} on the Weyl curvature and its derivative, working in the frame formulation to utilize the Cartan invariants. It is an improvement compared to proposal (\ref{prop}), which  in 5-dimensional  general relativity  works only for Schwarzschild black holes. Our proposal (\ref{entropyformula}) in 4 dimensions utilizes the Newman-Penrose scalar $\Psi_2$ (the only nonzero Weyl scalar for Petrov type D spacetimes), which is related to the Newman-Penrose spin coefficient\footnote{This is essentially the Bianchi identity, whose form holds because the tangent vector field for the geodesic null congruence is a double principal null direction of the Weyl tensor ($\Psi_0=\Psi_1=0$) for the systems we considered \cite[p.87]{stephani}.} $\rho$ by $D\Psi_2=3\rho \Psi_2$. The quantity  $\rho=||\nabla_a \tilde r||^2$ describes the divergence of the null geodesic congruence and therefore encodes the ``strength of gravitational field'' in focusing light rays, and hence also governs the evolution of area element of the horizon (as a trapped surface) \cite{callum}. Keeping in mind that entropy is a function of state: only changes of entropy between two different states are physically relevant, not the entropy of a given state.  We can therefore appreciate why it makes sense for $\Psi_2$ to appear in our gravitational entropy function, along with the frame derivative of the Weyl tensor which essentially encodes the effect of tidal forces. In the 5-dimensional case we cannot work with the Newman-Penrose $\rho$ or $\Psi_2$ because one cannot construct the null coframe. However, one can explicitly fix a frame and compute the Weyl tensor and frame derivatives with respect to this frame and work with those instead.

Much like the original proposal (\ref{prop}), we can only obtain the area law up to a constant prefactor. In addition, we have difficulties with deriving the entropy of black holes that is not of the standard $A/4$ form in some modified theories of gravity, such as $f(R)$ gravity and Einstein-Gauss-Bonnet gravity. However, this failure itself might be a hint at deeper physics. In order to obtain the correct form of the modified area law, in principle one has to choose other combinations of curvature quantities to construct a different entropy density function. Since many of these quantities have physical significance (e.g., Weyl curvature being related to the tidal deformation, $\rho$ being related to the convergence of light rays etc.), this suggests that gravitational entropy in different theories of gravity might be a manifestation of different physical effects. Consequently these black holes might have very different underlying microstates or microstructures -- whatever they might be -- from black holes in general relativity. 

While mathematically straightforward, a mystery remains regarding the \emph{physical} interpretation of taking the volume integral over $r\in (0,r_H)$, since for Schwarzschild black holes $r$ is a temporal coordinate; whereas for non-extremal Reissner-Nordstr\"om black holes $r$ is a temporal coordinate between the inner and outer horizons. (We note that the Clifton-Ellis-Tavakol proposal of gravitational entropy \cite{form7} when applied to a Schwarzschild black hole also integrates over $r$ in exactly the same manner.) On the other hand, in the case of cosmological spacetimes, the integral over $r$ out towards the apparent horizon is intuitively clear since $r$ is a radial coordinate. Why should these very different characters of $r$ be treated on equal footing? We leave this question open for future consideration.

We started off with the call for a further understanding of gravitational entropy with motivations stemming from the cosmological conundrum of the arrow of time, which emphasizes that in the very early Universe matter and radiation were in thermal equilibrium but the gravitational sector was not. It is perhaps curious as to why a proposal for gravitational entropy -- which we claim describes the entropy \emph{of} gravitational field -- would work to reproduce the entropy for a Reissner-Nordstr\"om black hole, which is a non-vacuum solution. Perhaps this has to do with the fact that gravitational entropy depends on the Weyl curvature, and Reissner-Nordstr\"om black hole belongs to Petrov class D along with Schwarzschild black hole -- that is, they share the same symmetry as far as the Weyl tensor is concerned. By the same logic, one would expect that the entropy of the Kerr solution can also be computed in a similar manner (see also, \cite{form7}).  By exploiting the Newman-Janis algorithm we showed that this is indeed the case if one takes into account the role played by the axis of rotation which breaks the spherical symmetry by picking a preferred spatial direction, though much work needs to be done in the future to fully understand the mathematical subtleties involved.

Bekenstein has introduced his entropy formula  as the Shannon {\it information entropy} applied to a black hole  \cite{bek}. On the other hand, Hawking has derived the same result invoking the concept of {\it thermodynamical entropy} by claiming that the black hole entropy should be related to the area because of the non-decreasing theorems \cite{texas}. In this paper, we have explicitly proved in some specific configurations that the Hawking-Bekenstein entropy can be regarded also as {\it gravitational entropy} because it can be written just in terms of the Weyl curvature without being sensitive to the matter content filling the exterior region in which the black hole is placed (indeed Bekenstein was interested in the entropy of the configurations inside the horizon \cite{bek1974} which is a crucial difference from the concept of {\it entanglement entropy} \cite{review}). As future projects we intend to investigate whether in cosmology our same formula can be interpreted also as {\it statistical entropy} (number of different inhomogeneous realizations on small scales which are compatible with the same coarse-grained large-scale homogeneous one) for tackling the structure formation problem; this would require one to check whether in an arbitrary spatial region the entropy is an increasing function of time in the same time intervals in which the spacetime shear is. For example, this latter quantity in the spacetime (\ref{generaldyn}) reads
\beq
\sigma^2= \frac{e^{-2 \nu(t,r)}[Y(t,r) \dot \lambda(t,r) -\dot Y(t,r)]^2}{3 Y^2(t,r)} \,.
\eeq 
It has already been argued that the formation of astrophysical structures could be tracked following the evolution of some of the entropy proposal we mentioned in Sect.\ref{ssintro} \cite{bolejko}, and in particular of the Kullback-Leibler relative {\it information entropy} \cite{buchert}; the latter however relies on the distribution of the cosmic matter density  whose direct connection to the Weyl curvature (e.g. tidal effects which trigger the gravitational collapse) is in general not self-evident.

\begin{acknowledgments}
DG is a member of the GNFM working group of the Italian INDAM. YCO thanks the National Natural Science Foundation of China (No.11922508) for funding support. The authors thank Shi-Qian Hu for discussions.
\end{acknowledgments}

{}

\end{document}